\documentclass{ws-procs9x6}
\usepackage{amssymb}
\usepackage{amsfonts}
\ifx\cal\undefined
    \def\cal #1{{\mathcal #1}}
\fi

\begin{document}

\title{SUPER-RADIANCE: \\FROM NUCLEAR PHYSICS TO PENTAQUARKS
\footnote{\uppercase{T}he collaboration with \uppercase{N}.
\uppercase{A}uerbach is highly appreciated. \uppercase{T}he
work was supported by the \uppercase{NSF} grant
\uppercase{PHY-0244453} and in part by
a grant from the \uppercase{US}-\uppercase{I}srael \uppercase{BSF}.}}

\author{Vladimir Zelevinsky}

\address{NSCL, Michigan State University, East Lansing, MI 48824-1321, USA}

\author{Alexander Volya}

\address{Department of Physics,
Florida State University, Tallahassee, FL 32306-4350, USA}

\maketitle

\abstracts{
The phenomenon of super-radiance in quantum optics predicted by
Dicke 50 years ago and observed experimentally has its
counterparts in many-body systems on the borderline between
discrete spectrum and continuum. The interaction of overlapping
resonances through the continuum leads to the redistribution of
widths and creation of broad super-radiant states and long-lived
compound states. We explain the physics of super-radiance and
discuss applications to weakly bound nuclei, giant resonances and
widths of exotic baryons.}

\section{Introduction}

Traditionally nuclear theory is divided into nuclear structure and
nuclear reactions. Being naturally related by physics of nuclei as
the subject of research, these fields are still significantly
different in their methods, quality and justification of
approximations and level of understanding. Moving away from the
line of nuclear stability, we have to overcome this barrier. Only
the consistent consideration of microscopic structure together
with the response of the system to external fields, including
various reaction amplitudes, is capable of developing the general
picture of loosely bound nuclei. Such systems which become {\sl
open} even under weak excitations have their specifics in extreme
sensitivity of internal properties to the proximity of continuum.
Similar problems emerge in atomic and molecular physics as well as
in mesoscopic condensed matter physics and quantum optics.

In a weakly bound quantum system, couplings of intrinsic dynamics
with reaction channels can crucially determine many important
physical properties. Above threshold, intrinsic energy levels
become resonances embedded in the continuum \cite{Mahaux} with the
lifetime typically shortening as excitation energy increases. The
relevant physical parameter here is $\kappa=\gamma/D$, the ratio
of the characteristic partial width (for a given channel) of a
resonance to the spacing between the resonances. The presence of
overlapping resonances, $\kappa>1$, is usually associated with a
very complicated and unpredictable pattern of Ericson fluctuations
that can only be considered in statistical terms of random
amplitudes. It turns out however that different, less known,
dynamics can take place in the region of a relatively small number
of open channels. New collective phenomena are possible leading to
the redistribution of the widths (and of corresponding time
scales) and creation of short-lived ({\sl super-radiant}) as well
as long-lived ({\sl trapped}) structures. Being a consequence of
the unitarity of the dynamics in the continuum, this restructuring
appears in any consistent theory that considers the bound states
and continuum on equal footing. In particular, it follows from
classical Fano theory \cite{Fano} widely used in quantum optics
\cite{Barnett}.

The term {\sl super-radiance} (SR) refers to the discovery by
Dicke \cite{Dicke} who 50 years ago has shown that, among $2^{N}$
states of a system of $N$ two-level atoms confined to the volume
of size smaller than the radiation wavelength between the two
levels, one state exists that radiates very fast and coherently so
that its width is close to $\Gamma=N\gamma$, where $\gamma$ is the
width of an individual isolated atom, and the intensity is
proportional to $N^{2}$. The remaining states live for a long time
having very small widths. The SR is observed \cite{SR} in the
laser pulse transmission through a resonant medium. It is
important that the coherent coupling of the atoms is reached due
to their interaction through the common radiation field,
independently of the direct atom-atom interaction.

The analog of the SR emerges in many-body quantum systems, Fig. 1,
where $N$ intrinsic states of the same symmetry are coupled to
common decay channels as was seen in numerical simulations for the
nuclear continuum shell model \cite{Kleinwa}. At $\kappa\sim 1$,
the system undergoes a phase transition from the separated narrow
resonances to the width accumulation by the SR states; their
number correlates with that of the open channels. The theory of
the phenomenon was given in \cite{SZ}, where the analogy to the
Dicke SR state was pointed out; for various aspects of theory see
\cite{Rotter,Sokann,Izrailev,VZ}.\\

Due to the very general character of the SR, it was observed in
many different situations in atomic, molecular, condensed matter
and nuclear physics. We present a brief review of the theory and
discuss selected applications to nuclear physics of low and
intermediate energies, including the suggestions for the SR as a
reason for the narrow width of exotic baryons.
\begin{figure}[ht]
\vskip -0.4 cm
\begin{minipage}[1]{2.4 in}
\centerline{\epsfxsize=2.2in\epsfbox{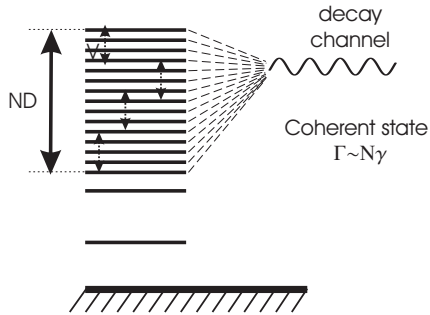}}
\end{minipage}
\begin{minipage}{1.8 in}
\caption{Schematic figure showing the SR mechanism. \label{fig1}}
\end{minipage}
\vskip -0.4 cm
\end{figure}

\section{Ingredients of the theory}

1. The convenient approach to the unified theory of intrinsic
states coupled to the continuum is provided by the Feshbach
projection techniques \cite{Feshbach}. The full system is
described by the Hermitian Hamiltonian $H$. The Hilbert space is
decomposed into two classes, internal, $Q\equiv\{|1\rangle\}$, and
external, $P=\{|c;E\rangle\}$, where $c$ marks the continuum
channels. The total eigenfunction at energy $E$ contains the
contributions of both classes; they are coupled by the matrix
elements $\langle c;E|H|1\rangle$. Eliminating the external states
by the projection {\sl at given energy}, one comes to the
effective eigenvalue problem in the intrinsic space, where the
effective Hamiltonian is given by
\begin{equation}
{\cal H}_{QQ}(E)=H_{QQ}+H_{QP}\,\frac{1}{E-H_{PP}+i0}\,H_{PQ}.
                                                \label{1}
\end{equation}

2. The internal energy-independent Hamiltonian $H_{QQ}$ determines
the eigenvalues $\epsilon_{n}$ of the states which would be bound
without the continuum coupling. This coupling converts at least
some of them into resonances with {\sl complex energies} ${\cal
E}_{j}=E_{j}-(i/2)\Gamma_{j}$.

3. The effective Hamiltonian (\ref{1}) depends on running energy
$E$. The continua $c$ are started at threshold energies $E_{c}$.
At $E>E_{c}$, the denominator of the propagator in eq. (\ref{1})
contains singular terms corresponding to the real ({\sl on-shell})
decay into the channel $c$ with energy conservation. The real part
$\Delta$ of the propagator corresponds to the principal value of
the singular terms and describes the virtual ({\sl off-shell})
processes of coupling through all (closed and open) channels,
while only open channels contribute to the imaginary part
$(-i/2)W$ that makes the effective Hamiltonian {\sl
non-Hermitian}.

4. The anti-Hermitian part of the effective Hamiltonian is
factorized into a number of terms equal to the number $k$ of
channels open at energy $E$. Thus, the general form of ${\cal H}$
is
\begin{equation}
{\cal H}_{12}=H_{12}+\Delta_{12}(E)-\frac{i}{2}\,W_{12}(E), \quad
W_{12}=\sum_{c;\,{\rm open}}A^{c}_{1}A^{c\ast}_{2}, \label{2}
\end{equation}
where the amplitudes $A^{c}_{1}$ for the coupling of the intrinsic
state $|1\rangle$ to the channel $c$ are proportional to the
original matrix elements $\langle 1|H|c\rangle$.

5. The principal part $\Delta$ renormalizes the intrinsic
Hamiltonian $H_{QQ}$. In many situations it is of minor
significance being small and weakly dependent on energy. The
anti-Hermitian part is the driving force for new physics. The
amplitudes $A_{1}^{c}$ should vanish at threshold energy $E_{c}$,
and this energy dependence is crucial for weakly bound systems.

6. The diagonalization of ${\cal H}$, eq. (\ref{2}), determines
the complex eigenvalues ${\cal E}_{j}=\tilde{E}_{j}-
(i/2)\Gamma_{j}$ and the (biorthogonal) set of the eigenfunctions
$|j\rangle$ of quasistationary states depending on running energy
$E$. The resonance centroid $E_{j}$ on the real axis is
self-consistently determined by $\tilde{E}_{j}(E=E_{j})=E_{j}$.

7. The same formalism determines the reaction cross sections at
a given energy with the aid of the scattering matrix
\begin{equation}
S^{ab}(E)=s_{a}^{1/2}\left\{\delta^{ab}-\sum_{12}A^{a\ast}_{1}\,
\left(\frac{1}{E-{\cal H}}\right)_{12}A^{b}_{2}\right\}
s_{b}^{1/2}.                                       \label{3}
\end{equation}
Here the ``potential" phases $s_{a}=\exp(2i\delta_{a}(E))$
describe the contributions of remote resonances which usually are
not taken into account explicitly. The full effective Hamiltonian
${\cal H}$, eq. (\ref{2}) in the denominator in eq. (\ref{3}) includes
the same amplitudes $A_{1}^{c}$ as in the numerator, and this
makes the $S$-matrix explicitly unitary.

\section{SR phase transition}

At energy below the lowest threshold, we have the Hermitian
Hamiltonian $H+\Delta$. This determines the discrete spectrum of
bound states $\epsilon_{n}$ and the stationary wave functions
$|\psi_{n}\rangle$. The unitary transformation to this basis of
the {\sl internal representation} keeps the anti-Hermitian part
factorized so that at energy in the continuum
\begin{equation}
{\cal H}_{nn'}=\epsilon_{n}\delta_{nn'}-\frac{i}{2}W_{nn'}, \quad
W_{nn'}=\sum_{c;\,{\rm open}}A_{n}^{c}A^{c\ast}_{n'}. \label{4}
\end{equation}
The strength of the continuum coupling in channel $c$ is measured
by the ratio $\kappa_{c}=\gamma_{c}/D$ of the typical partial
width $\gamma_{c}=|A^{c}|^{2}$ to the mean level spacing $D$ (we
consider the class of intrinsic states with the same exact quantum
numbers). At small $\kappa_{c}$, we have in this channel isolated
narrow resonances with the widths $\sim \gamma_{c}$.

As $\kappa_{c}$ increases, the role of the {\sl off-diagonal
damping} \cite{Barnett} increases. When $\kappa_{c}$ is
getting close to 1, the {\sl percolation} happens: decay of a
level $n$ with return to the overlapping level $n'$ becomes likely,
and all levels are coherently coupled through the continuum,
similarly to the Dicke coherent state. Now we are close to another
limit where the dynamics are defined by $W_{nn'}$. At strong
coupling one can start with the {\sl doorway representation} of
the eigenstates of $W$. Due to the factorization of $W$, the rank
of this matrix is equal to the number $k$ of open channels, and
there are $k$ doorway states with non-zero widths. The remaining
$N-k$ states are weakly coupled to the continuum only through the
doorways. The eigenstates of ${\cal H}$ are divided into two
groups: the SR states sharing almost all summed width of the $N$
states, and the trapped states which are very long-lived.

In the simplest limit of a single open channel and degenerate
intrinsic spectrum, $\epsilon_{n}\equiv \epsilon$, the whole width
$\Gamma={\rm Tr}\,W$ is concentrated in one SR state. If the
intrinsic levels $\epsilon_{n}$ are not degenerate but their
spread $\delta\epsilon\sim ND\ll \Gamma$, the sum $\tilde{\Gamma}$
of all small $N-1$ widths is \cite{AZDelta}
\begin{equation}
\tilde{\Gamma}\approx 4\,\frac{(\Delta\epsilon)^{2}}{\Gamma}.
                                             \label{5}
\end{equation}

The continuum coupling in the limit of $\kappa\sim (\Gamma/ND)\gg
1$ implements the segregation of distinct time scales in the
reaction process:

the shortest time $\tau_{d}\sim\hbar/\Gamma$ corresponds to the
fast direct reaction through the SR intermediate state;

fragmentation time $\tau_{f}\sim\hbar/\Delta\epsilon\sim
\kappa\tau_{d}$ is what is necessary for the internal damping of
the original excitation of one of the intrinsic states;

Weisskopf time $\tau_{W}\sim\hbar/D\sim N\kappa \tau_{d}$ is the
recurrence time of the wave packet (intrinsic equilibration);

compound lifetime of trapped states $\tau_{c}\sim\hbar/
(\tilde{\Gamma}/N)\sim\kappa \tau_{W}$ that allows for the full
exploration of intrinsic space.

\section{Some applications}

The universality of the SR mechanism guarantees its appearance in
any situation with a not very large number of open channels
(otherwise the off-diagonal damping is quenched by the Ericson
fluctuations of amplitudes corresponding to different channels)
and the necessary ingredients present. There are many examples of
the manifestation of the SR in molecular, atomic and condensed
matter physics, see for instance \cite{Verevkin,Flam,Rotter2}.
Below we shortly discuss some examples from nuclear physics.

\subsection{Two collectivities}.
Giant resonances (GR) in the response function of the nucleus to
the multipole excitation appear as a result of the coherent
coupling of simple particle-hole excitations. By our terminology,
they are generated by the intrinsic interaction that accumulates
the {\sl multipole strength} at some energy shifted from the
interval $\Delta\epsilon$ of the unperturbed intrinsic states. The
strength collectivization, however, is not what is seen in the
reaction. The observed cross sections are determined by the
partial {\sl widths} of unstable intrinsic states with respect to
a given channel.

The simple model
\cite{Zagreb,Sokrot} with the effective Hamiltonian
\begin{equation}
{\cal H}_{nn'}=\epsilon_{n}\delta_{nn'} +\lambda d_{n}d_{n'}
-\frac{i}{2}A_{n}A_{n'},                      \label{6}
\end{equation}
contains unperturbed energies, real multipole interaction and
interaction through the continuum. The dynamics here are
determined by two multidimensional vectors ${\bf d}=\{d_{n}\}$ and
${\bf A}=\{A_{n}\}$. The multipole interaction creates the GR
shifted by $\Omega=\lambda {\bf d}^{2}$ {\sl along the real energy
axis} from the unperturbed centroid $\bar{\epsilon}$ and
accumulating the multipole strength. The continuum interaction
creates the SR state shifted by $\Gamma={\bf A}^{2}$ {\sl along
the imaginary energy axis} to the lower part of the complex plane
and accumulating almost all available decay width.

The interplay of the collective effects depends on the ``angle"
$\phi$ between the vectors ${\bf d}$ and ${\bf A}$. In the
degenerate case, $\epsilon_{n}=\epsilon$, the reaction amplitude
looks like
\begin{equation}
T(E)=\frac{(E-\epsilon-\Omega)\Gamma+\lambda ({\bf A}\cdot{\bf
d})^{2}}{(E-\epsilon-\Omega)[E-\epsilon+(i/2)\Gamma]+(i/2)\lambda
({\bf A}\cdot{\bf d})^{2}}.                    \label{7}
\end{equation}
In the limits of parallel internal and external couplings,
$\phi=0^\circ$, we have
\begin{equation}
T(E)=\frac{\Gamma}{E-\epsilon-\Omega +(i/2)\Gamma}. \label{8}
\end{equation}
Here the SR and GR collectivization are combined, and the
experiment will reveal the shifted ``{\sl Giant Dicke resonance}"
with full strength and full width. This is what we expect to see
in gamma scattering, where both intrinsic excitation and decay
have the same multipole nature. In the opposite case of orthogonal
couplings, $\phi=90^\circ$, the result is
$T(E)=\Gamma/[E-\epsilon-(i/2)\Gamma].       $
Now the experiment would show only the unshifted SR resonance with
vanishing multipole strength but broad width (the decay channel
has another nature, for example results from evaporation). The GR
state is {\sl dark} with collective strength but no access to the
continuum. Fig. 2 shows a more realistic case of non-degenerate
intrinsic states when both the strength and the width are shared
between the displaced GR and the SR in the region of unperturbed
excitation energies. As the coupling with continuum increases, the
low-energy branch is expected to develop into what is called {\sl
pigmy giant resonance}. We hope to present the study of the pigmy
resonance elsewhere but the mechanism shown here is quite
universal. Because of complicated interference between the two
types of collectivity, the observed picture should differ in
different channels.
\begin{figure}[ht]
\vskip -0.5 cm
\centerline{\epsfxsize=4.0in\epsfbox{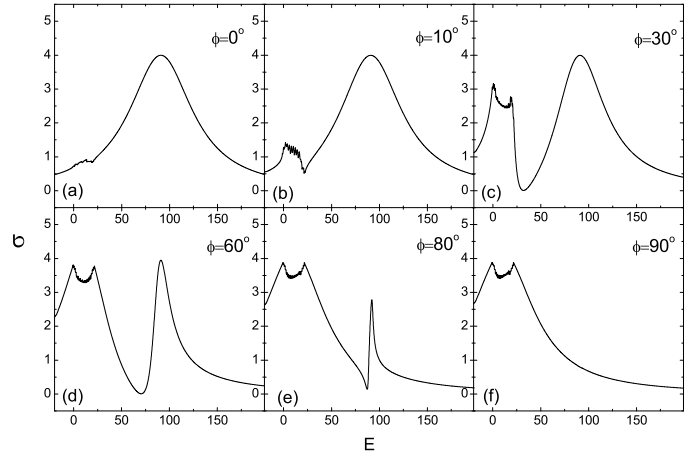}} 
\vskip -0.2 in
\caption{The
scattering amplitude in the model of eq. (6) with 20 states with
spacing between $\epsilon_n$ set as a unit of energy; the
parameters are $\Omega=\Gamma=80$. The panels show the evolution
of scattering as a function of increasing angle between the
multidimensional vectors ${\bf d}$ and ${\bf A}$.\label{fig2}}
\vskip -0.3 in
\end{figure}

\subsection{Loosely bound nuclei}

\begin{figure}[ht]
\vskip -0.3 cm 
\centerline{\epsfxsize=4.0in\epsfbox{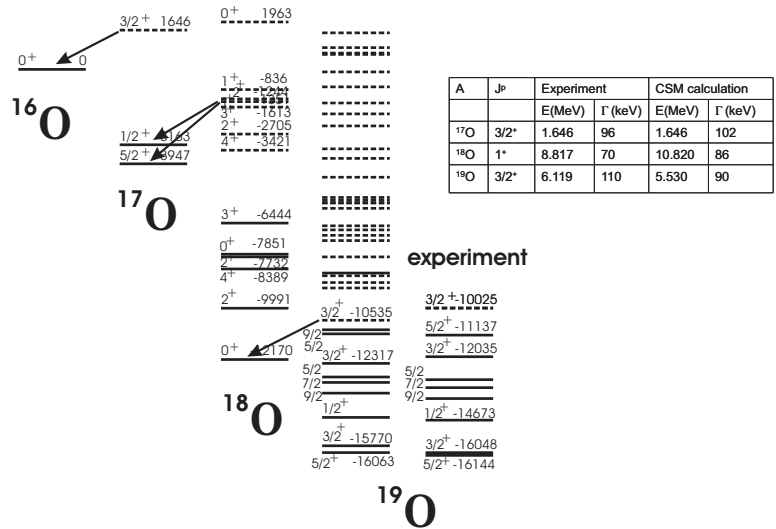}}
\caption{\label{fig3} Continuum shell model calculation for the
oxygen isotopes with $A=16$ to 19. One and two-neutron emission
processes are considered. On a single energy scale stable nuclear
states are shown with solid lines and neutron-unstable states by
dashed lines. The experimentally observed states of $^{19}$O are
shown to the right on the same energy scale. The scarce
experimental data for energies and neutron decay widths are listed
in the table along with theory predictions. Correspondingly, in
the level scheme the arrows indicate the decay transitions of
these states. }
\vskip -0.3 in
\end{figure}

These applications stimulated the renewed interest to  the SR
physics. As we have discussed at the previous Seminar \cite{Aldo},
in the proximity of thresholds the coupling of intrinsic states
with and through the continuum dominates the dynamics. The shell
model machinery can be generalized \cite{CSM} to include the
effective non-Hermitian energy-dependent Hamiltonian. In
particular, one can consistently consider systems like $^{11}$Li
where the single-particle states are unbound but additional
correlations bring paired states back from the continuum. In all
such cases the correct energy dependence of the amplitudes
$A_{1}^{c}$ near threshold is crucial. However, this dependence is
sensitive to the absolute position of thresholds. To solve the
problem in a nucleus $A$ with one- and two-body decay channels, we
need to know the spectrum of the $(A-1)$ and $(A-2)$ nuclei, and
the whole chain of daughter nuclei. The plausibility of such a
program was demonstrated for the chain of oxygen isotopes
\cite{CSM}. Fig. \ref{fig3} shows full shell model calculation for
the few oxygen nuclei above the $^{16}$O core and comparison with
data; this will be further discussed in a forthcoming publication.

\section{Widths of exotic baryons}
The SR mechanism works in hadron physics. A baryon resonance $R$
created, for example, in a photonuclear experiment mixes with the
$RN^{-1}$ states of $R$-(nucleon hole) nature. The baryon
resonances have excitation energies starting from few hundred MeV
and typical vacuum widths exceeding 100 MeV. The continuum mixing
with the $RN^{-1}$ states of the same symmetry produces an SR
state and few trapped states which are observed as narrow
resonances on the broad SR background. This pattern was seen in
the $^{12}$C$(e,e'p\pi^{-}$)$^{11}$C Mainz experiment \cite{Mainz}
at the $\Delta$-isobar region and attributed \cite{AZDelta} to the
SR mechanism. In terms of two collectivities, the SR state here is
a good candidate for the parallel case. As shown by old RPA
calculations \cite{Hirata}, the pionic coherent state (similar to
the nuclear GR) accumulates the pion decay width as well.
\begin{figure}[ht]
\vskip -0.5 cm
\begin{minipage}[1]{2.2 in}
\epsfxsize=2.2in\epsfbox{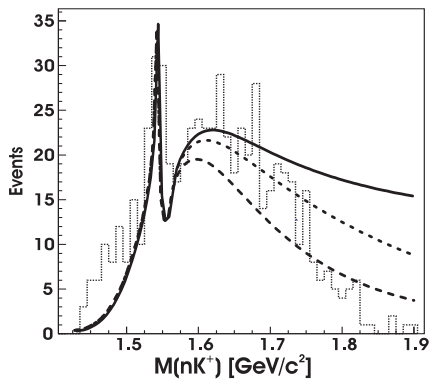}
\end{minipage}
\hskip 0.1 in
\begin{minipage}[1]{2.1 in}
\caption{\label{fig4}The two-state model for the pentaquark in
comparison with the results of the CLAS experiment $\gamma
\,d\rightarrow p\,K^{-}(K^{+}\,n)$ $^{24}$. The $p$-wave
$(K^{+}\,n)$ continuum channel is considered with kinematical
dependence $A(E)\sim E^{3/4}$ near threshold, solid curve. For two
other curves, the amplitude $A(E)$ at high energy is damped as
$A(E)\sim [E^3/(1+E/\Lambda)^3]^{1/4}$ and the cutoff parameters
$\Lambda$ are 300 MeV, dashed curve, and 500 MeV, dotted curve.}
\end{minipage}
\vskip -0.2 in
\end{figure}

We have applied \cite{AZVTheta} the same idea to recent
observations by different groups of the $\Theta^{+}$ resonance
with strangeness +1. The resonance coined as {\sl pentaquark} has
a very narrow width smaller than few MeV. In the experiments on
nuclei the situation should be quite similar to that in the
$\Delta$-case. In \cite{AZVTheta} the arguments are given that in
the deuteron and proton experiments the same mechanism can still
be at work because the intrinsic QCD dynamics create few states of
different nature but the same symmetry which are coupled through
the continuum. It is sufficient to have just two such states, for
example (but not necessarily) a standard quark bag and a large
size quasimolecular state. Fig. 4 shows the result for a simple
two-state model with naturally chosen parameter values where the
narrow resonance on the broad background emerges due to the SR
mechanism.

\section{Conclusion}

The phenomenon of SR is very general; it has to appear, provided
all necessary ingredients are in place, in any theory that
describes physics on the borderline of discrete and continuum
spectrum and agrees with the unitarity requirements. The
segregation of physical time scales is accompanied by the
appearance of long-lived states on a broad background. One can
expect the most spectacular manifestations in many-body systems
(autoionizing atomic states, chemical reactions, loosely bound
nuclei, collective dynamics in the continuum and analogous physics
on a quark level). There are important ramifications for quantum
chaos and its experimental tests, mesoscopic condensed matter
physics, studies of entanglement, decoherence and quantum
information.


\begin{thebibliography}{99}

\bibitem{Mahaux} C. Mahaux and H.A. Weidenm\"{u}ller, {\sl Shell
Model Approach to Nuclear Reactions} (North Holland, Amsterdam,
1969).

\bibitem{Fano} U. Fano, Nuovo Cim. {\bf 12} (1935) 156; Phys. Rev.
{\bf 124} (1961) 1866.

\bibitem{Barnett} S.M. Barnett and P.M. Radmore, {\sl Methods in
Theoretical Quantum Optics} (Clarendon Press, Oxford, 2002).

\bibitem{Dicke} R.H. Dicke, Phys. Rev. {\bf 93} (1054) 99.

\bibitem{SR} N. Scribanowitz, I.P. Herman, J.C. MacGillivray, and
M.S. Fields, Phys. Rev. Lett. {\bf 30} (1973) 309.

\bibitem{Kleinwa} P. Kleinw\"{a}chter and I. Rotter, Phys. Rev. C
{\bf 32} (1985) 1742.

\bibitem{SZ} V.V. Sokolov and V.G. Zelevinsky, Phys. Lett. B {\bf
202} (1988) 10; Nucl. Phys. {\bf A504} (1989) 562.

\bibitem{Rotter} I. Rotter, Rep. Prog. Phys. {\bf 54} (19910 635.

\bibitem{Sokann} V.V. Sokolov and V.G. Zelevinsky, Ann. Phys.
(N.Y.) {\bf 216} (1992) 323.

\bibitem{Izrailev} F.M. Izrailev, D. Saher, and V.V. Sokolov,
Phys. Rev. E {\bf 49} (1994) 130.

\bibitem{VZ} A. Volya and V. Zelevinsky, J. Opt. B {\bf 5} (2003)
450.

\bibitem{Feshbach} H. Feshbach, Ann. Phys. (N.Y.) {\bf 5} (1958)
357; {\bf 19} (1962) 287.

\bibitem{AZDelta} N. Auerbach and V. Zelevinsky, Phys. Rev. C {\bf
65} (2002) 034601.

\bibitem{Verevkin} V.B. Pavlov-Verevkin, Phys. Lett. A {\bf 129}
(1988) 168.

\bibitem{Flam} V.V. Flambaum, A.A. Gribakina, and G.F. Gribakin,
Phys. Rev. A {\bf 54} (1996) 2066.

\bibitem{Rotter2} I. Rotter, E. Persson, K. Pichugin, and P. Seba,
Phys. Rev. E {\bf 62} (2000) 450.

\bibitem{Zagreb} V.V. Sokolov and V.G. Zelevinsky, Physika
(Zagreb) {\bf 22} (1990) 303.

\bibitem{Sokrot} V.V. Sokolov, I. Rotter, D.V. Savin, and M.
M\"{u}ller, Phys. Rev. C {\bf 56} (1997) 1031, 1044.

\bibitem{Aldo} V. Zelevinsky and A. Volya, Challenges of Nuclear
Structure, Proceedings of the 7th International Spring Seminar on
Nuclear Physics, ed. A. Covello (World Scientific, Singapore,
2002) p. 261.

\bibitem{CSM} A. Volya and V. Zelevinsky, Phys. Rev. C {\bf 67}
(2003) 054322.

\bibitem{Mainz} P. Bartsch {\sl et al.}, Eur. Phys. J. A {\bf 4}
(1999) 209.

\bibitem{Hirata} M. Hirata, J.H. Koch, F. Lenz, and E.J. Moniz,
Ann. Phys. (N.Y.) {\bf 120} (1979) 205.

\bibitem{AZVTheta} N. Auerbach, V. Zelevinsky, and A. Volya. Phys.
Lett. B {\bf 590} (2004) 45.

\bibitem{Step} S. Stepanyan {\sl et al.} (CLAS collaboration),
Phys. Rev. Lett. {\bf 91} (2003) 252001.

\end{thebibliography}
\end{document}